\documentclass[aps,prd,superscriptaddress,floatfix,nofootinbib,preprintnumbers,eqsecnum,twocolumn]{revtex4-1}

\pdfoutput=1
\usepackage{amssymb,amsfonts,amsmath,graphicx}
\usepackage{xcolor}
\usepackage{orcidlink}
\usepackage[english]{babel}
\usepackage[T1]{fontenc}
\usepackage{amsmath}
\usepackage{slashed}
\usepackage{booktabs}
\usepackage{listings}
\usepackage[utf8]{inputenc}
\usepackage{hyperref}
 \hypersetup{ 
   colorlinks,
   linkcolor={blue!80!black},
   citecolor={blue!70!black},
   urlcolor={blue!70!black}
 }
 
\usepackage{placeins}

\usepackage{amssymb,amsmath,epsfig,todonotes,longtable}

\numberwithin{equation}{section}

\newcommand{\bean}{\begin{eqnarray*}}
\newcommand{\eean}{\end{eqnarray*}}

\newcommand{\fref}[1]{Figure~\ref{#1}}

\newcommand{\cN}{{\cal N}}

\newcommand{\cA}{{\cal A}}
\newcommand{\cB}{{\cal B}}
\newcommand{\cC}{{\cal C}}

\newcommand{\cV}{{\cal V}}

\def\cjn1{{\cA, \cC^*\otimes \wedge^j \cN^*}}
\def\bjn1{{\cA, \cB^*\otimes \wedge^j \cN^*}}
\def\vjn1{{\cA, \cV^*\otimes \wedge^j \cN^*}}
\def\cjn2{{\cA, \cC\otimes \wedge^j \cN^*}}
\def\bjn2{{\cA, \cB\otimes \wedge^j \cN^*}}
\def\vjn2{{\cA, \cV\otimes \wedge^j \cN^*}}

\newcommand{\be}{\begin{equation}}
\newcommand{\ee}{\end{equation}}
\newcommand*{\nnbe}{\begin{equation}}
\newcommand*{\nnee}{\end{equation}}
\newcommand{\bea}{\begin{eqnarray}}
\newcommand{\eea}{\end{eqnarray}}
\newcommand{\ba}{\begin{align}}
\newcommand{\ea}{\end{align}}
\newcommand{\bi}{\begin{itemize}}
\newcommand{\ei}{\end{itemize}}

\newsavebox{\overlongequation}

\begin{document}
\title{Cosmic Inflation and Genetic Algorithms}



\author{Steve A.~Abel \orcidlink{0000-0003-1213-907X}}
\email[]{s.a.abel@durham.ac.uk}
\affiliation{IPPP, Durham University, Durham, DH1 3LE UK\\ Department of Mathematical Sciences, Durham University,
Durham DH1 3LE, UK}

\author{Andrei Constantin \orcidlink{0000-0002-0861-5363}}
\email[]{andrei.constantin@physics.ox.ac.uk}
\affiliation{Rudolf Peierls Centre for Theoretical Physics, University of Oxford, Parks Road, Oxford OX1 3PU, UK}

\author{Thomas R.~Harvey \orcidlink{0000-0002-4990-4778}}
\email[]{thomas.harvey@physics.ox.ac.uk}
\affiliation{Rudolf Peierls Centre for Theoretical Physics, University of Oxford, Parks Road, Oxford OX1 3PU, UK}

\author{Andre Lukas \orcidlink{0000-0003-4969-0447}}
\email[]{andre.lukas@physics.ox.ac.uk}
\affiliation{Rudolf Peierls Centre for Theoretical Physics, University of Oxford, Parks Road, Oxford OX1 3PU, UK}


\begin{abstract}
Large classes of standard single-field slow-roll inflationary models consistent with the required number of e-folds, the current bounds on the spectral index of scalar perturbations, the tensor-to-scalar ratio, and the scale of inflation can be efficiently constructed using genetic algorithms. The setup is modular and can be easily adapted to include further phenomenological constraints. A semi-comprehensive search for sextic polynomial potentials results in $\sim\mathcal{O}(300,000)$ viable models for inflation. The analysis of this dataset reveals a preference for models with a tensor-to-scalar ratio in the range $0.0001\leq r\leq0.0004$. We also consider potentials that involve cosine and exponential terms. In the last part we explore more complex methods of search relying on reinforcement learning and genetic programming. While reinforcement learning proves more difficult to use in this context, the genetic programming approach has the potential to uncover a multitude of viable inflationary models with new functional forms. 
\end{abstract}

\maketitle		

\section{Introduction}
The idea of cosmic inflation, a period of exponential growth for the scale factor of the Universe believed to have taken place between $10^{-36}$ and $10^{-32}$ seconds after the Big Bang singularity, was originally introduced 
to explain the initial conditions for the hot Big Bang model, which would otherwise lead to a number of conundrums such as the horizon problem, the flatness problem and the monopole (and other relics) problem~\cite{Starobinsky:1980te, Guth:1980zm, Linde:1981mu, Albrecht:1982wi, Linde:1983gd}. Moreover, the theory provides a plausible account for the origin of structure in the Universe and explains the observed cosmic microwave background anisotropies by providing a mechanism for the generation of density perturbations~\cite{Mukhanov:1981xt, Hawking:1982cz, Guth:1982ec, Starobinsky:1982ee, Bardeen:1983qw}.

The inflationary epoch is typically modelled as a quasi-exponential expansion of the very early universe caused by the slow-roll of a scalar field $\phi$, called the inflaton, in its evolution towards the minimum of a nearly flat potential $V(\phi)$. In this slow-roll regime, all the derivatives of the scalar field are negligible in comparison with the potential $V(\phi)$, interpreted as an approximately constant vacuum energy. The effective energy density and pressure of a homogeneous scalar field $\phi(t)$ are
\begin{equation}\label{eq:phi_state}
\rho_\phi = \frac{1}{2}\dot\phi^2 + V(\phi)~,\qquad p_\phi = \frac{1}{2}\dot\phi^2 - V(\phi)~,
\end{equation}
hence in the slow-roll regime $p_\phi\approx-\rho_\phi$, approximating the equation of state associated with a cosmological constant term. 
The expression for $V(\phi)$ should in principle be derived from a fundamental theory such as string theory, however this task entails myriad subtleties and difficulties. This paper will take a phenomenological approach therefore, and consider certain, suitably parametrised classes of functions $V(\phi)$.

The concrete properties of the inflaton potential that yield a viable phenomenology are that it must vanish at its global minimum and should have a nearly flat region near the minimum to induce the standard slow-roll dynamics. Within this region, the starting point for inflation must be sufficiently far from the global minimum to generate a large enough expansion in order to solve the cosmological problems of the Big Bang model. The minimum required amount of inflation is about $70$ e-folds, which corresponds to an expansion by a factor of $10^{30}$. All estimates of the minimal number of e-folds suffer from uncertainties of up to $10$ e-folds, depending on the details of reheating. 

More stringent constraints on the inflationary potential come from measurements of the CMB anisotropies which give information about the primordial density perturbations and gravitational waves. The results included in the 2018 Planck cosmological release~\cite{Planck:2018jri} are compatible with a simple power law spectrum for the scalar perturbations with an exponent $n_s-1$ and a measured value of the spectral index $n_s = 0.9649\pm 0.0042$ at the pivot scale $k_\ast=0.05~{\rm Mpc}^{-1}$. The collaboration has also set tight constraints on the amount of inflationary gravitational waves with an upper limit on the tensor-to-scalar ratio $r<0.061$. The bound on $r$ can be rephrased as an upper bound on the energy scale of inflation $V_\ast$ when the pivot scale exited the Hubble radius given by $V_\ast< (1.6\cdot 10^{16}~{\rm GeV})^4$ \cite{Planck:2018jri}. However, the energy scale of inflation is fixed by the normalisation of the present-day power spectrum of density perturbations $(V_\ast/\epsilon_\ast)^{1/4}=6.7\cdot 10^{16}~{\rm GeV}$ with a theoretical uncertainty of about $10\%$ \cite{Bunn:1996py}, where $\epsilon$ is one of the slow-roll parameters. In general, inflationary models predict a wide range of values for the number of e-folds, $n_s$, $r$ and the energy scale of inflation, and many of them can be ruled out on this basis. 

Finding inflationary potentials, possibly falling within some class  that is motivated by an underlying theory, that satisfy all these conditions is a nontrivial task. For example, it involves performing an integral to determine the number of e-folds. Such a task is not amenable to direct optimisation techniques such as gradient descent, simulated annealing, or more direct forms of machine learning. It is not a simple regression problem and it is not, for example, easy to express the problem directly in a single loss-function. Moreover finding good solutions is made more difficult by the fact that functions may have ``local optima'', namely regions that satisfy some but not all the conditions for inflationary potentials, which may be widely separated from good solutions by large ``barren'' regions. The solving of such convoluted systems is much more naturally expressed as the optimisation of a ``fitness'' within some ``environment'', which is comprised of the set of operations that must be performed on the proposed function in order to tell if it is a good solution or not. This suggests using either reinforcement learning (RL) (for a review see~\cite{sutton2018reinforcement}) or 
genetic algorithms (GAs)~\cite{turing,Holland1975,David1989,Holland1992,Forrest1993,Jones1995,Collard1998,Reeves2002,haupt,Michalewicz2004} as a means of finding solutions. 

The  purpose of this paper is to show that large classes of viable inflationary potentials can indeed be constructed using GAs.  Although GAs have been used, somewhat sporadically, in the context of particle physics and string theory~\cite{Yamaguchi:1999hq,Allanach:2004my,Akrami:2009hp,Blaback:2013ht,Abel:2014xta,Abel:2018ekz,Cole:2019enn,AbdusSalam:2020ywo,Bena:2021wyr,Abel:2021rrj,Abel:2021ddu,Cole:2021nnt,Loges:2021hvn}, to our knowledge, they have not yet been used for the purpose of building cosmological models. In this paper we will be focusing on this specific goal of determining viable inflationary potentials. We will consider potentials whose functional dependence includes polynomials, exponentials and trigonometric functions as building blocks. 

Within each class of functions, parametrised by a set of coefficients, we will search for those coefficient values that lead to viable inflationary models. The numerical ranges for the coefficients are discretised, turning the search into a discrete optimisation problem that is amenable to heuristic methods such as GAs or RL. 
GAs prove to be extremely efficient in identifying many viable potentials, while we find that RL turns out to be less successful. Thus GAs appear to provide a useful tool for inflationary model building. 

Although GAs are the main focus of this paper, it is natural to ask whether more sophisticated forms of symbolic regression could be employed for the same purpose. In the final section we comment on the possible implementation of more elaborate algorithms that incorporate the evolutionary version of symbolic regression, namely genetic programming (GP). This could in principle simultaneously optimise for the functional dependence of the potential and its numerical coefficients. We present arguments suggesting that GP should be effective for the task of constructing simple inflationary potentials using a set of allowed operands if it is implemented in conjunction with the GA using {\it speciation}, that is the GP works at the level of a population of functions, while the GA is used to determine the coefficients in each function by optimising its fitness.  

\section{Genetic search algorithm}
\subsection{Binary encoding of inflationary potentials}
In this study we will concentrate on three different classes of potentials: (1)~polynomial, (2) polynomial + $\cos(c\phi)$, and (3) polynomial + $e^{c\phi}\times$ polynomial, where all the polynomials are chosen to have a fixed degree. Potentials of this type have been previously studied in the literature in various theoretical contexts.  

The polynomial coefficients, as well as the numerical factor~$c$ in the argument of the exponential or the cosine represent the parameters of the model. In the study that we present in Section~\ref{sec:results}, we will specify certain ranges of possible values for each parameter and then discretise them. The set of all the discrete parameter values forms our space of states $\{s\}$. The polynomial coefficients must be allowed to vary over several orders of magnitude and, accordingly, their discretisation will be spaced on a logarithmic scale. In view of the genetic algorithm implementation to be discussed below, the number of partitions is always chosen to be a power of $2$ such that, for any given parameter, each discrete value can be assigned a binary label. Finally, all the labels are concatenated together into one long binary string that encodes the inflationary model. All the models belonging to the same functional class are encoded on strings of equal lengths. 

\subsection{The genetic algorithm}
The starting point of a GA is the creation of an initial random population with a certain number, $N_{\rm pop}$, of individuals. This is done by generating $N_{\rm pop}$ binary strings of the appropriate length, referred to as the {\it genotype}, each representing an inflationary potential, referred to as the {\it phenotype}, in the class of functions under consideration. This population is then evolved using the three main ingredients specific to GAs: selection, breeding and mutation. The optimal size of the population is correlated with the length of the binary string and in our case $N_{\rm pop}$ will be of order of a few hundred individuals.

Selection proceeds by ranking the individuals according to their fitness value. In our case the fitness value will be a measure of how compatible a given model of inflation is with the current cosmological observables. The detailed form of the fitness function is discussed below. Subsequently, individuals are selected for breeding with a probability that depends linearly on their ranking by fitness, such that the probability of the $k^{\rm th}$ individual being selected for breeding is 
\begin{equation}
P_k~=~ \frac{2}{(1+\alpha)N_{\rm pop}} \left( 1+\frac{N_{\rm pop}-k}{N_{\rm pop}-1}(\alpha-1)\right)~.
\end{equation}
In particular, the probability  $P_1$ for the top individual is equal to a multiple $\alpha$ of the probability $P_{N_{\rm pop}}$ of the least fit individual.
Typically, $\alpha$ is chosen in the range $2\leq\alpha\leq 5$. While the fittest individuals have a higher chance to produce offsprings in the next generation, the scheme also allows less fit individuals to breed. This ensures a variety of ``genes" is available throughout the evolution. 

Breeding is implemented as an $M$-point cross-over by which the two binary sequences are cut at the same $M$ random points and the cut sections are swapped in an alternating fashion. In our case a single point cross-over performs well enough and we will consider this simple choice. Finally, once a new generation is formed, a small fraction of the binary digits, selected randomly, are flipped, to ensure that the population does not stagnate. A typical value for the mutation rate is one percent. 

The population is evolved for multiple generations, typically a few hundred. If successful, the algorithm produces a final population where a large fraction of the individuals correspond to viable solutions. The search is then restarted with a new random population. As we will see, this form of genetic algorithms is capable of quickly identifying many high-quality solutions within each evolutionary cycle, even though the fraction of viable solutions is typically very small compared to the size of the state space. This remarkable property of GAs leads to a significant reduction of computing time, as compared to systematic scanning procedures. 

In the present context it is worth mentioning a possible but inadvisable modification, because the reason for its failure is instructive. It is tempting when dealing with a set of discretised parameters to divide the genotype into sections (``chromosomes" perhaps) with one section corresponding to each parameter. Then one could implement a cross-over by a one-point cross-over between {\it all} the corresponding chromosomes of the two individuals. However, this is exactly the wrong thing to do in a GA: when the parameters are arranged into one long string with only a single cross over, during breeding individuals will be swapping several parameters wholesale as well as mixing up the single parameter that contains the cross-over. Such swapping of large chunks of genotype is an important aspect of GAs. If one attempts to split the genome into sections that represent parameters, one is in a sense allowing the phenotype to ``constrict" the genotype and this greatly diminishes the efficacy of the GA.

\subsection{The fitness value for inflationary models}\label{sec:FitnessValue}
There are several ways in which a potential can fail to be a viable model of inflation. By analysing these failures it is possible to come up with a sensible definition for the fitness function required for the GA implementation. We will adopt the view that the potential $V(\phi)$ is an effective field theory description for the dynamics of $\phi$ and that its expression can only be trusted in a relatively small range $I_\phi$ of inflaton values, which we choose to be the interval $I_{\phi}=[-30,30]$ in Planck units. (We will be working in units where the reduced Planck mass $1/\sqrt{8\pi G_N}$ has been set to one.)

The potential must have a minimum within this range. Moreover, as the entire inflationary trajectory must fall within~$I_{\phi}$, we require that the position of the lowest minimum $\phi_m$ within~$I_{\phi}$ (which from now on will be referred to as the ``global" minimum) falls inside a smaller interval $I_{\phi_m}$, which we choose as $I_{\phi_m}=[-5,5]$. When generating potentials, we do not require the vacuum value at the global minimum to vanish but simply adjust the additive constant in $V$ (which is not considered part of the state space) such that $V(\phi_m)=0$. If $V$ has no minima inside $I_{\phi_m}$, the fitness value of the model is set to a fixed negative value (``no-vacuum penalty").

Once the global minimum $\phi_m$ has been found, we identify the closest maxima to the right and to the left of $\phi_m$, since the inflaton can in principle roll down from either side. The regions between the global minimum and the nearest maxima are of interest for inflation. If no maximum is found within $I_\phi$ (to the right or to the left), the region of interest is extended up to the corresponding boundary of $I_\phi$. The fitness function is computed for both the left and the right trajectories, and the trajectory with the larger fitness is adopted as the relevant one for the given potential.

To discuss our algorithm further we need to introduce the notion of the slow-roll approximation; which is characterised by the smallness, $\epsilon\ll1$ and $\eta\ll 1$, of the slow-roll parameters
\begin{equation}\label{SR}
\epsilon ~=~ \frac{1}{2} \left(\frac{V'}{V}\right)^2\;,\qquad \eta ~=~   \frac{V''}{V}\;.
\end{equation}
In general, when the slow-roll conditions are satisfied inflation occurs. We begin our algorithmic exploration of the potential close to the global minimum $\phi_m$ where $V$ is small and hence the slow-roll conditions are violated. Moving away from the global minimum in sufficiently small discrete steps $\Delta\phi$ we seek the value of $\phi$ closest to $\phi_m$ where the slow-roll parameters drop  below a certain threshold, for concreteness taken to be
\begin{equation}\label{SRcond}
\epsilon\leq 0.5\;,\qquad \eta\leq 0.5\; .
\end{equation}
The corresponding value $\phi_e$ provides the approximate field value which marks the end of inflation.

To find the starting point of inflation, we continue to move up the potential in steps $\Delta\phi$ and continue to check the slow-roll conditions. At the same time, in order to avoid the production of new inflationary domains which expand at faster rate (eternal inflation), we require that the quantum fluctuations of $\phi$ during a typical time interval $H^{-1}$ are much smaller than the change in the field due to its classical motion, that is
\begin{equation}\label{QF}
\delta \phi_q ~\approx~\frac{\sqrt{V/3}}{2\pi}~\ll~ 2\pi\frac{V'}{3V} ~=~ \delta \phi_c~.
\end{equation}
In practice, we check that $\delta\phi_q<0.5\, \delta \phi_c$. As long as the conditions in Eqs.~\eqref{SRcond} and \eqref{QF} are satisfied, the standard slow-roll inflation approximations can be trusted. The field value $\phi_i$ where inflation starts is then either the value closest to $\phi_e$ where any of the conditions in Eqs.~\eqref{SRcond} and \eqref{QF} fails to hold or, if the conditions hold all the way to the boundary, the corresponding end point of $I_\phi$. 

Having obtained the values of the field corresponding to the start and end of inflation, we proceed to the computation of the cosmological observables. First, we compute the total number of e-folds, 
\begin{equation}
N_{\rm total} ~=~ \int_{\phi_e}^{\phi_i} \frac{V}{V'} d\phi~.
\end{equation}
This number should be large enough to solve the standard cosmological problems of the Big Bang model. More importantly, it should be larger than the number of e-folds separating the end of inflation $\phi_e$ from the reference value $\phi_\ast$ when the cosmological scales of interest (that is, the scales probed by the Planck satellite) leave the horizon. The useful formula to compute this number of e-folds (assuming the reheating scale is not too far away from $\sim 10^{10} $GeV)  is~\cite{Lyth:1998xn} 
\begin{equation}
N_\ast ~\approx~ 58 +\frac{1}{6} \ln V(\phi_e)~,
\end{equation}
which can be computed as soon as the field value $\phi_e$ at the end of inflation is known. On the other hand,
\begin{equation}
N_{\ast} ~=~ \int_{\phi_e}^{\phi_\ast} \frac{V}{V'} d\phi~,
\end{equation}
from which the value of the field $\phi_\ast$ at the pivot scale can be obtained.  

For an acceptable model of inflation we require $N_{\rm total}\geq N_\ast$. In order to penalise a model which violates this condition, the fitness functions receives a contribution
\begin{equation}\label{fitnessN}
f_N(s)~=~-\frac{1}{10}(N_\ast-N_{\rm total})\, \theta(N_*-N_{\rm total})\; ,
\end{equation} 
where $\theta(x)$ is the Heaviside function. Having obtained the value of the field $\phi_\ast$ at the pivot scale, we can compute the cosmological observables 
\begin{equation}
n_{s,\ast} ~=~ 1-6\epsilon_\ast + 2\eta_\ast~,\qquad r_\ast ~=~ 16\epsilon_\ast~.
\end{equation}

To enforce the correct spectral index, the fitness function includes the term 
\begin{equation}\label{fitnessns}
f_{n_s}(s)~=~\hat{f}_{n_s}(s)\theta(f_{\rm term} - \hat{f}_{n_s}(s))~,
\end{equation} 
where
\begin{equation}
\hat{f}_{n_s}(s)~=~-\log_{10}\left(1+\frac{|n_{s,\ast} - n_s|}{\Delta n_s} \right)~,
\end{equation} 
and the measured value is $n_s\pm \Delta n_s = 0.9649\pm 0.0042$. Here, $f_{\rm term}$ is a fixed value (see below) such that models with $\hat{f}_{n_s}(s)>f_{\rm term}$ attract a vanishing fitness contribution and are considered leading to a viable spectral index.

If the tensor-to-scalar ratio $r_\ast$ is greater than the maximal valued $r_{\rm max} = 0.061$ consistent with the 2018 Planck release, the fitness receives a further penalty of
\begin{equation}\label{fitnessr}
f_r(s)~=~-\log_{10}\left( \frac{r_\ast}{r_{\rm max}}\right)\,\theta(r_*-r_{\rm max})\; .
\end{equation} 

Finally, we compute $m_\ast=(V_*/\epsilon_*)^{1/4}$ and to ensure that the model matches the energy scale of inflation $m\simeq 0.027$ (in Planck units) we add the term
\begin{equation}\label{fitnessm}
f_m(s)~=~ \hat{f}_{m}(s)\theta(f_{term} - \hat{f}_{m}(s))
\end{equation}
where
\begin{equation}
\hat{f}_m(s)~=~ -\log_{10}\left(1+\frac{|m_\ast-m|}{\Delta m}\right)
\end{equation}
to the fitness function, where we have set $\Delta m = 0.1\, m$ to account for theoretical uncertainties.

The fitness  of a state $s$ is then given by the sum of the above contributions, that is, $f(s)=f_N(s)+f_{n_s}(s)+f_r(s)+f_m(s)$.
We declare a state $s$ viable, or `terminal', if $f(s)=0$, where we choose $f_{\rm term}=-0.3$. This fitness function is linked to the GA Mathematica package {\tt Genetic} which was developed by the authors and which is available for download~\cite{GApackage}. The results discussed in the following section have been obtained within this set-up.

\section{Results}\label{sec:results}
\subsection{Polynomial potentials}
One of the simplest classes of functions that can be considered are polynomials. These can arise in inflationary models in their own right or as truncated Taylor expansions around the global minimum of more general perturbative potentials. For concreteness, we restrict to polynomials of degree six,
\begin{equation}
	V(\phi) ~=~ c_0+c_1 \phi + c_2 \phi^2 + c_3 \phi^3 + c_4 \phi^4 + c_5 \phi^5 + c_6 \phi^6  .
\end{equation}
The constant term $c_0$ is fixed by requiring that $V(\phi)$ vanishes at the global minimum, as discussed in Section~\ref{sec:FitnessValue}. The GA search is then set up to identify optimal values for the remaining six constants $c_1,c_2,\ldots, c_6$, which run over the following ranges: 
\begin{equation}
\begin{tabular}{c|c|c}
~~coefficient~~ &~~ range~~ &~~ no.~of~partitions~~\\\hline\hline
$c_1$ & $~~(-10^{-8},10^{-8})~~$ & $64$\\
$c_2$ & $~~(-10^{-9},10^{-9})~~$ & $64$\\
$c_3$ & $~~(-10^{-10},10^{-10})~~$ & $64$\\
$c_4,c_5,c_6$ & $~~(-10^{-11},10^{-11})~~$ & ~$64$ \vspace{-0.4cm}
\end{tabular}\vspace{-0.1cm}
\end{equation}\\
Each of these intervals is discretised into $64$ values, equally spaced on a logarithmic scale, leading to a length $6$ binary string for each coefficient and a binary string of length $36$ for each state. Hence, the state space consists of a total of $2^{36}\simeq 7\cdot 10^{10}$ models.
\begin{figure}[!h]%
    \hspace{-0.2cm}
    \includegraphics[width=8.7cm]{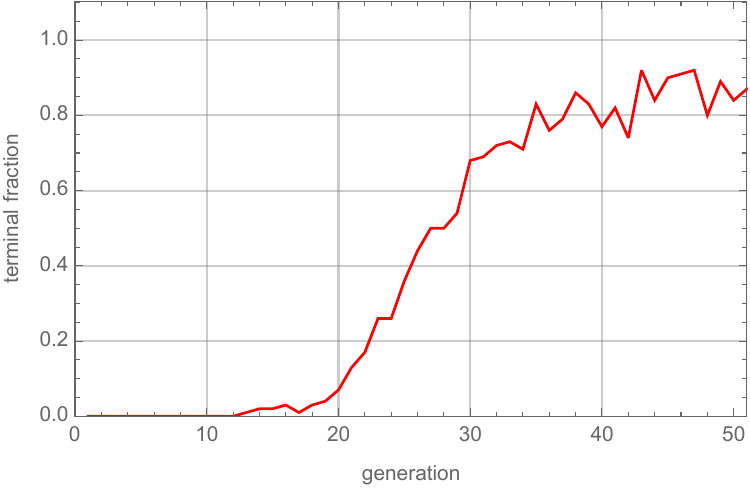} %
    \caption{\small Evolution of the fraction of viable models.}%
\label{fig:FractionPolyn}
\end{figure}
The GA is run for a population of $N_{\rm pop}=100$ individuals, a mutation rate of $1\%$ and a coefficient $\alpha=3$. A typical evolution over $50$ generations leads to ${\cal O}(10)$ different viable models. This is extremely efficient, given that only about $5000$ models are visited in the course of such an evolution and, by comparison, a typical random selection of $5000$ models would contain no viable potentials at all. Figure~\ref{fig:FractionPolyn} shows the evolution of the fraction of terminal models in the course of a typical run, which takes about $200\,{\rm s}$ to complete on a standard laptop. 

\begin{figure}[!h]%
    \centering
    \includegraphics[width=8.cm]{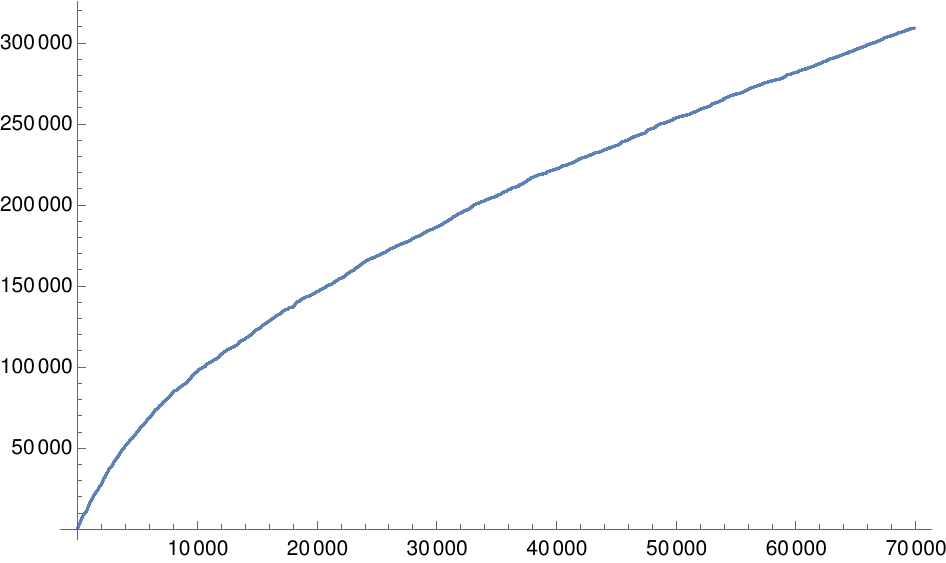} %
    \caption{\small Number of distinct viable models found vs.~the number of iterations of GA. Each iteration consisting of 50 generations with a population of 100.}%
\label{fig:SaturationPolyn}
\end{figure}

To obtain a more comprehensive dataset of viable polynomial potentials of degree six, we have performed $70,000$ runs, each starting from a random initial population. The total number of visited states is $\sim 10^8$, which amounts to  $\sim 1\%$ of the search space. The search produced $309,097$ different viable models. Figure~\ref{fig:SaturationPolyn} shows a plot of the number of distinct viable models (that is, the number of viable models after eliminating repetitions from previous runs) as a function of the number of visited states. The curve indicates a progression towards saturation, suggesting that a sizeable fraction of the viable models have been found, although even for this simple model it was not possible to reach complete saturation.\\

We illustrate the discussion with two examples. The first example corresponds to the potential 
\begin{equation}
\begin{aligned}
V(\phi) & ~=~ 3.57328\cdot10^{-10}+\,1.77313\cdot10^{-10} \phi \\
&\,\,\,\,\,\,\,\,\,\,\, +\, 1.77313\cdot10^{-11}\, \phi^2 - 1.13662\cdot10^{-12}\, \phi^3  \\
&\,\,\,\,\,\,\,\,\,\,\, +\, 1.45\cdot10^{-16}\, \phi^4 + 1.95409\cdot10^{-15}\, \phi^5  \\
&\,\,\,\,\,\,\,\,\,\,\, -\,   1.0\cdot10^{-16}\, \phi^6~,\label{eq:polyex1}
 \end{aligned}
\end{equation}\\
\begin{figure}[!h]%
    \hspace{-0.2cm}
    \includegraphics[width=8.7cm]{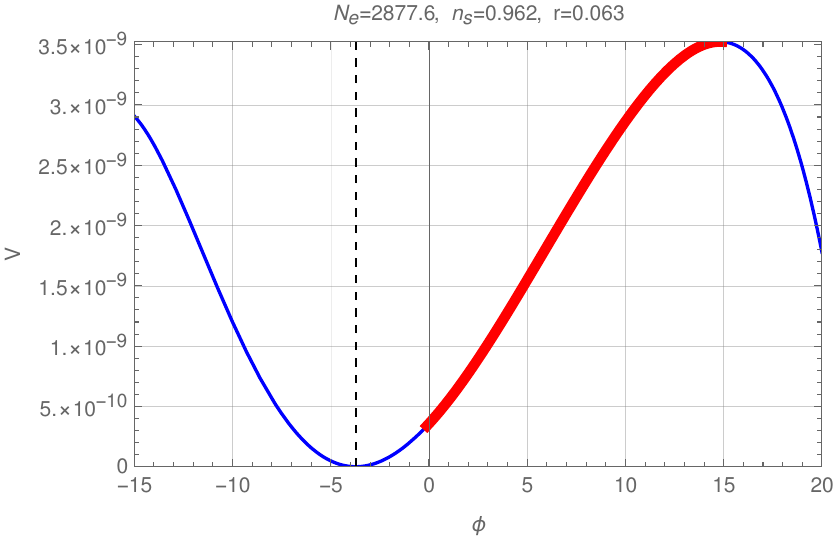} %
    \caption{\small The polynomial potential of degree 6 from Eq.~\eqref{eq:polyex1}: large field inflation.}%
\label{fig:PolynEx1}
\end{figure}
\noindent \hspace{-0.18cm}which is shown in figure~\ref{fig:PolynEx1}. For the chosen example, the inflation region is shown in red in Figure~\ref{fig:PolynEx1}, starting at a value $\phi_i \simeq 14.8$ and ending at $\phi_e\simeq 0$. The model generates a total number $N=2900$ of e-folds. The minimum of the potential is at $\phi_m\simeq -3.7$. The value of the field at the pivot scale is $\phi_\ast \simeq 9.4$ with the corresponding number of e-folds $N_\ast \simeq 53.4$ until the end of inflation. The scale of inflation, the spectral index and the tensor-to-scalar ratio are given by $m=(V_\ast/\epsilon_\ast)^{1/4}\simeq 0.029$, $n_{s,\ast} \simeq 0.9620$ and $r_\ast\simeq 0.063$, respectively. With the total field excursion approximately equal to $|\phi_*-\phi_e|\simeq 10$ in Planck units, this is an example of large field inflation. 

This particular model appears with an effective seven-fold degeneracy without our search, corresponding to a cluster of "nearby" models which only differ by tiny changes in the parameters. In principle the search can be refined to remove such effective degeneracies in parameter space. In terms of the GA a modification to remove such degeneracy could also be implemented directly using a crowding penalty (where the crowding in question is in the phenotype). Unfortunately such a modification would be very costly in search time: for each new solution  one would have to compute its distance to all of the $ \mathcal O(10^5)$ previous solutions. Therefore we present the raw counts, with the understanding that the number of non-degenerate solutions may be an order of magnitude fewer than the raw number. In addition we do not divide out for solutions that are related by a displacement in $\phi$. A potentially much more efficient method for separating the results into independent solutions would be post-processing, first by separating them into bins with different numbers of e-folds and $r_
\ast$, and then using a clustering algorithm on the bins individually. We will not attempt this in the present study.\\

\begin{figure}[!h]%
    \hspace{-0.2cm}
    \includegraphics[width=8.7cm]{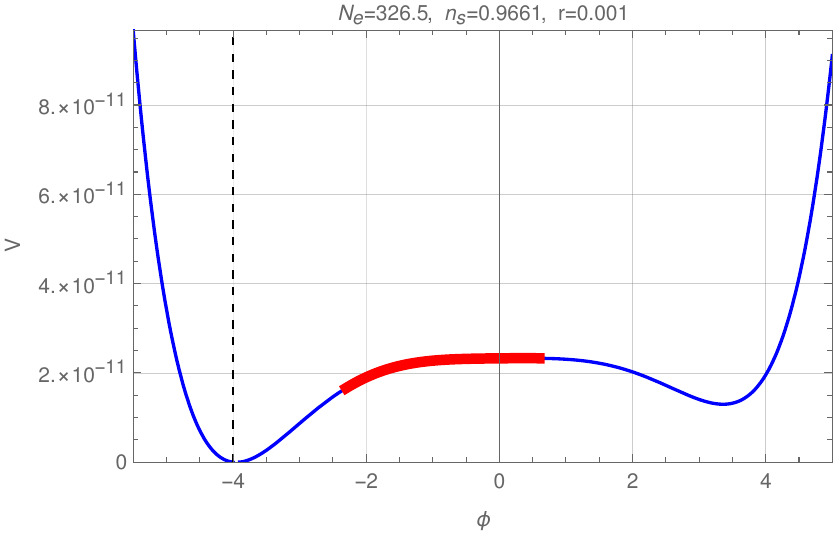} %
    \caption{\small The polynomial potential of degree 6 from Eq.~\eqref{eq:polyex2}: small field inflation.}%
\label{fig:PolynEx2}
\end{figure}

\vspace{12pt}
The second example we will present is the potential 
\begin{align}
V(\phi) & ~=~ 2.32562\cdot 10^{-11} + 1.44128\cdot10^{-13} \phi \nonumber \\ 
&\,\,\,\,\,\,\,\,\,\,\,  -\, 1.33159\cdot10^{-13}\, \phi^2 + 5.47169\cdot10^{-15}\, \phi^3\nonumber\\
 &\,\,\,\,\,\,\,\,\,\,\,  -\, 2.44753\cdot10^{-13}\, \phi^4 + 8.63806\cdot10^{-15}\, \phi^5 \nonumber \\
 &\,\,\,\,\,\,\,\,\,\,\, +\, 1.25252\cdot10^{-14}\, \phi^6\; , \label{eq:polyex2}
 \end{align}
which has a global minimum at $\phi_m\simeq -4.0$. The inflation region (shown in red in Figure~\ref{fig:PolynEx2}) starts at $\phi_i \simeq 0.5$ and ends at $\phi_e\simeq -2.2$, hence this is an example of small field inflation. The total number of e-folds allowed by this model is $N\simeq 330$ and the pivot scale is $\phi_\ast \simeq -0.2$, corresponding to $N_\ast \simeq 53.9$ e-folds until the end of inflation. The scale of inflation, the spectral index and the tensor-to-scalar ratio are $m=(V_\ast/\epsilon_\ast)^{1/4}\simeq 0.028$, $n_{s,\ast} \simeq 0.9661$ and $r_\ast\simeq 0.001$, respectively.

\subsection{Statistics on the class of polynomial potentials}
\begin{figure}[!h]%
    \centering
    \includegraphics[width=8.5cm]{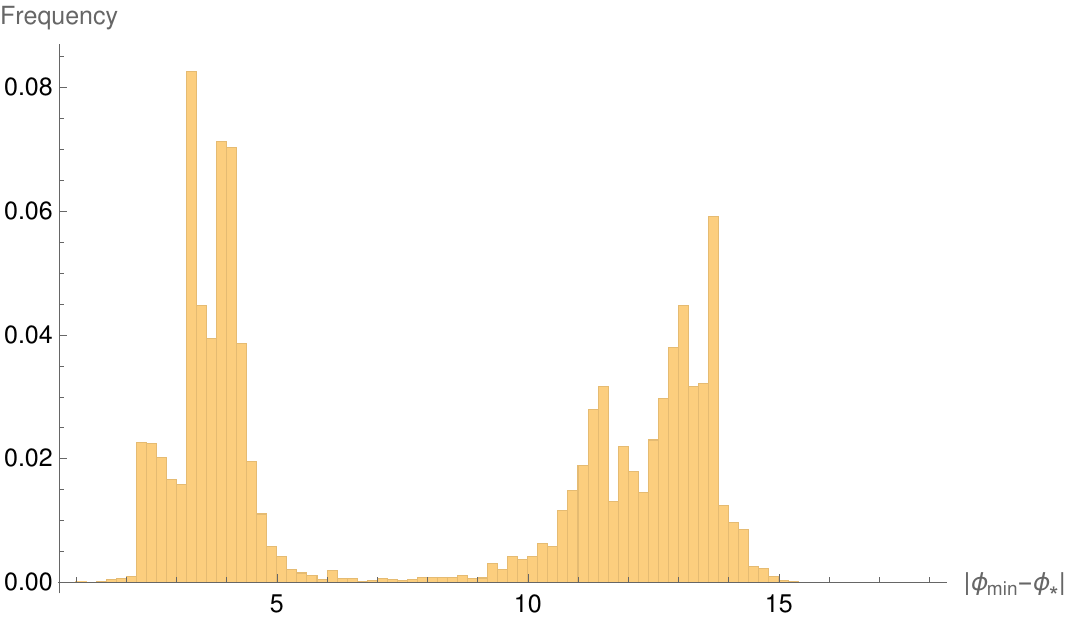} %
    \caption{\small Minimum field excursions during inflation, normalised as a probability distribution (i.e. the sum of all the bars in the histogram is unity).}%
\label{fig:FieldRangeStats}
\end{figure}

\begin{figure}[!h]%
    \centering
    \includegraphics[width=8.5cm]{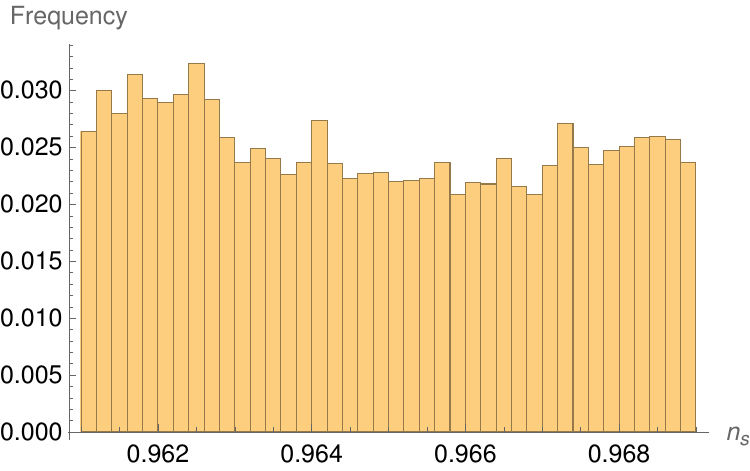} %
    \caption{\small Histogram for the spectral index $n_s$.}%
\label{fig:NsStats}
\end{figure}

\begin{figure}[!h]%
    \centering
    \includegraphics[width=8.5cm]{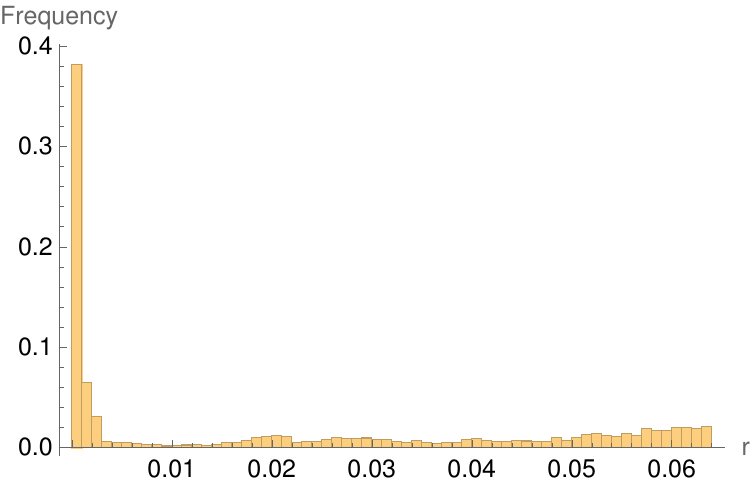} %
    \caption{\small Histogram for the tensor-to-scalar ratio $r_\ast$.}%
\label{fig:rStats}
\end{figure}

\begin{figure}[!h]%
    \centering
    \includegraphics[width=8.5cm]{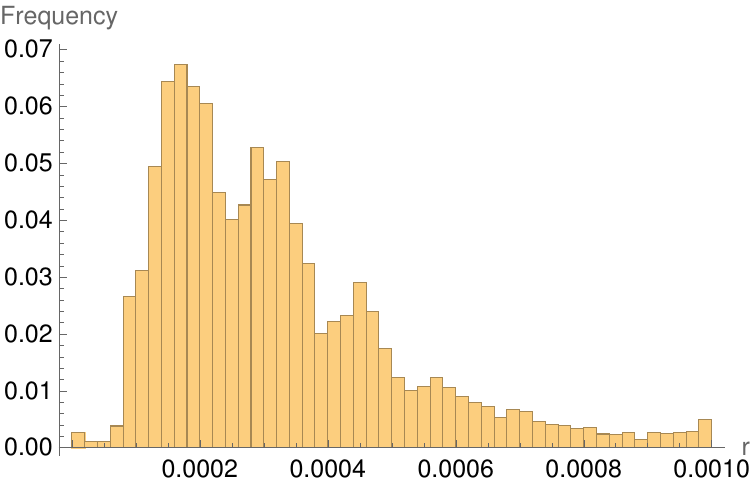} %
    \caption{\small Zoomed-in tensor-to-scalar ratio prediction, $r_\ast$.}%
\label{fig:rStats3}
\end{figure}

The large number of viable inflationary potentials found during the GA search $(\sim0.3\cdot 10^6)$ allows us to generate some meaningful statistics. We can first look at the field range during which the universe expands by $N_\ast$ e-folds before the end of inflation. The histogram for the minimum field excursion $|\phi_\ast-\phi_m|$ during inflation~\footnote{Of course, inflation can continue for larger field excursions. However, this is the minimum range required  to satisfy all constraints.} is displayed in Figure~\ref{fig:FieldRangeStats} and shows two peaks, one corresponding to small field inflation (an $\sim {\cal O}(1)$ field excursion in Planck units) and another one corresponding to large field inflation (an $\sim {\cal O}(10)$ field excursion). It is interesting to note that a very small fraction of the models exhibit mid-range field excursions or excursions larger than 15 Planck units. 
The distribution for the spectral index $n_s$, shown in Figure~\ref{fig:NsStats}, is approximately uniform, indicating no particular preference shown by $n_s$ within the allowed range. On the other hand, the distribution for the tensor-to-scalar ratio $r_\ast$, shown in Figure~\ref{fig:rStats}, indicates a strong preference towards models with a small value $r_\ast <0.002$. Zooming in on the $r$-histogram (see Figure~\ref{fig:rStats3}), we note that the most favoured value of $r_\ast$ lies in the range $0.0001\leq r_\ast \leq0.0004$, which can be regarded as a prediction for the value of the tensor-to-scalar ratio and a hint for why primordial gravitational waves have not yet been detected. 


\FloatBarrier

\subsection{Polynomial + Cosine}
Potentials involving a cosine function appear in what is known as `natural inflation' in which the inflaton field is a
pseudo-Nambu Goldstone boson appearing after the breaking of a shift symmetry (see for example Ref.~\cite{Martin:2013tda} for a review of the main types of inflation proposed over the years and their theoretical justification). Here we use a generalisation of this type of potential which is of the form 
\begin{equation}
	V(\phi) \,=\, c_0+c_1 \phi + c_2 \phi^2 + c_3 \phi^3 + c_4 \phi^4 + c_5\cos(c_6 \phi) .
\end{equation}
As in the previous case, the constant term $c_0$ is fixed by requiring that $V(\phi)$ vanishes at the global minimum. The  ranges of search for the remaining six parameters $c_1,c_2\ldots, c_6$ are chosen as follows: 
\begin{equation}
\begin{tabular}{c|c|c}
~~coefficient~~ &~~ range~~ &~~ no.~of~partitions~~\\\hline\hline
$c_1$ & $~~(-10^{-8},10^{-8})~~$ & $64$\\
$c_2$ & $~~(-10^{-9},10^{-9})~~$ & $64$\\
$c_3$ & $~~(-10^{-10},10^{-10})~~$ & $64$\\
$c_4,c_5$ & $~~(-10^{-11},10^{-11})~~$ & $64$\\
$c_6$ & $~~(0,5)~~$ & ~$64$ \vspace{-0.4cm}
\end{tabular}\vspace{-0.1cm}
\end{equation}\\
The intervals for $c_1,c_2,c_3,c_4$ and $c_5$ are again divided into $64$ values, equally spaced  logarithmically. For $c_6$, which enters the argument of the cosine, the interval is divided in $64$ values, equally spaced. As before, an inflaton potential is, therefore, described by a binary list of length $36$ and the state space consists of $2^{36}\simeq 7\cdot 10^{10}$ models. With the same GA settings as above, the computational time for a single run is comparable to that of the  sextic polynomial case. A typical run results in ${\cal O}(10)$ different viable models. 

\begin{figure}[!h]%
    \hspace{-0.2cm}
    \includegraphics[width=8.7cm]{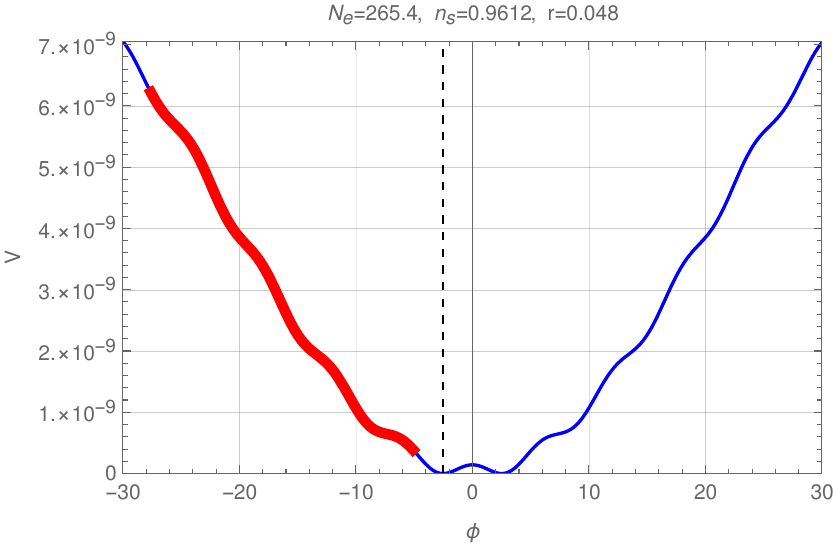} %
    \caption{\small The polynomial + cosine potential from Eq.~\eqref{cosex}: large field inflation.}%
\label{fig:PolynCos}
\end{figure}

A typical example in this class is shown in Figure~\ref{fig:PolynCos} and corresponds to the potential
\begin{align}
\label{cosex}
V(\phi) & \,=\, 2.80939\cdot10^{-11} + 5.92241\cdot10^{-14} \phi\\
& \,\,\,\,\,\,\,\, +\, 1.13662\cdot10^{-11} \phi^2 - 3.79642\cdot10^{-16} \phi^3\nonumber \\
& \,\,\,\,\,\,\,\, -\, 4.10847\cdot10^{-15} \phi^4 + 1.3662\cdot10^{-10} \cos(1.05 \phi)\; , \nonumber
 \end{align}
which has a global minimum at $\phi_m\simeq -2.5$, a pivot at $\phi_\ast \simeq -13.2$ and the end of slow roll at $\phi_e\simeq -5.2$. Hence this is an example of large field inflation. The total number of e-folds allowed, in the range we consider, by this model is $N\simeq 265$. At the pivot scale the corresponding number of e-folds $N_\ast \simeq 54.4$, with the scale of inflation, the spectral index and the tensor-to scalar ratio given by $m=(V_\ast/\epsilon_\ast)^{1/4}\simeq 0.028$,  $n_{s,\ast}\simeq 0.9611$ and $r_\ast\simeq 0.048$, respectively.

\FloatBarrier 

\subsection{Polynomial + Exp $\times$ polynomial}
Finally we consider inflationary potentials motivated by K\"ahler moduli inflation in string theory. The functional form considered is 
\begin{align}
V(\phi)& \,\,=\,\, c_0+c_1 \phi + c_2 \phi^2 + c_3 \phi^3 + c_4 \phi^4 \\
&\,\,\,\,\,\, + (c_6+c_7 \phi + c_8 \phi^2 + c_9 \phi^3 + c_{10} \phi^4)\,e^{c_5 \phi}~.\nonumber
\end{align}
As before, the constant term $c_0$ is fixed by requiring that $V(\phi)$ vanishes at the global minimum. The  search ranges for the remaining ten parameters is chosen as follows: 
\begin{equation}
\begin{tabular}{c|c|c}
~~coefficient~~ &~~ range~~ &~~ no.~of~partitions~~\\\hline\hline
$c_1,c_7$ & $~~(-10^{-8},10^{-8})~~$ & $64$\\
$c_2,c_8$ & $~~(-10^{-9},10^{-9})~~$ & $64$\\
$c_3,c_9$ & $~~(-10^{-10},10^{-10})~~$ & $64$\\
$c_4,c_6,c_{10}$ & $~~(-10^{-11},10^{-11})~~$ & $64$\\
$c_5$ & $~~(-12.4,12.4)~~$ &  ~$64$ \vspace{-0.4cm}
\end{tabular}\vspace{-0.1cm}
\end{equation}\\
For all the coefficients except $c_5$ the intervals  are divided into $64$ values, equally spaced logarithmically, while $64$ equally spaced values are used for $c_5$. Hence, each state is described by a binary string of length $60$ and the total size of the state space is $2^{60}\simeq 10^{18}$. This increase in the size of the search space, compared to the previous cases, does not pose a problem for the GA. 
With the same GA settings as before, $N_{\rm pop}=100$ individuals, a mutation rate of $1\%$ and a coefficient $\alpha=3$, a typical evolution over $50$ generations leads to ${\cal O}(1000)$ different viable models and completes within a few minutes on a standard laptop. Compared to the pure sextic polynomial case, viable potentials are easier to find in this setting. A modest number of 1000 runs leads to over a million distinct viable potentials. 

We present two viable models from this class. The first corresponds to small field inflation and is given by the potential 
\begin{align}
V(\phi) ~=~ & \, 5.26402\cdot10^{-11} + 1.56\cdot10^{-14} \phi \nonumber \\
&\hspace{-0.2cm} + 10^{-15} \phi^2 + 3.79642\cdot10^{-16} \phi^3\nonumber\\
&\hspace{-0.2cm} - 2.1025\cdot10^{-16} \phi^4 \nonumber\\
&\hspace{-0.2cm}+  e^{1.2\, \phi} \big{(}-6.9349\cdot10^{-12} \nonumber\\
&\hspace{-0.2cm} - 1.33159\cdot10^{-13} \phi^2\nonumber\\ 
&\hspace{-0.2cm} + 10^{-16} \phi^3 + 1.68795\cdot10^{-13} \phi^4\big{)}~.\label{exex}
\end{align}
\begin{figure}[!h]%
    \hspace{-0.2cm}
    \includegraphics[width=8.7cm]{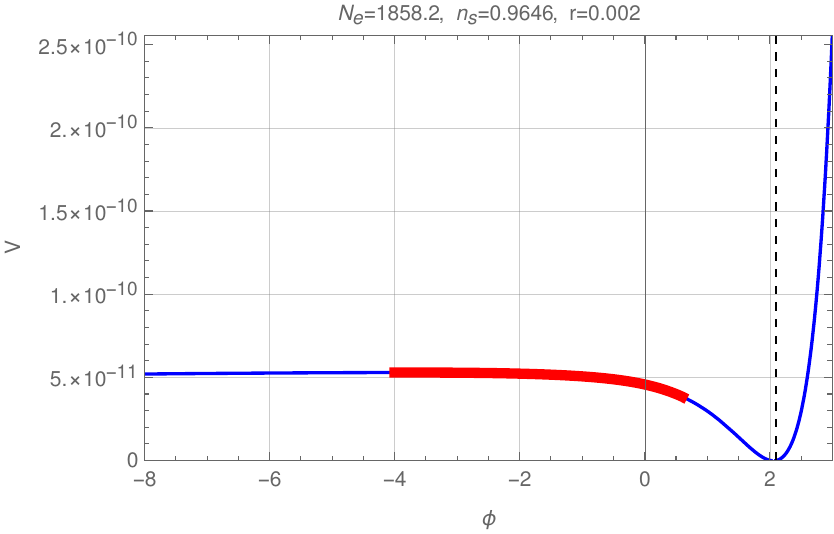} %
    \caption{\small The quartic + exp $\times$ quartic potential from Eq.~\eqref{exex}: small field inflation.}%
\label{fig:PolynExp2Ex1}
\end{figure}

The potential is shown in~\fref{fig:PolynExp2Ex1}. The global minimum at $\phi_m\simeq 2.1$ is crucially determined by the exponential term. Slow roll starts at a value $\phi_i \simeq  -3.8$ and ends at $\phi_e\simeq 0.5$. The value of the field at the pivot scale is $\phi_\ast \simeq -2.1$, corresponding to  $N_\ast \simeq  53.9$ e-folds until the end of inflation. Hence, this can be considered a small-field inflationary model. The maximum number of e-folds allowed in this model is $N\simeq 1860$. The scale of inflation, the spectral index and the tensor-to-scalar ratio are $m=(V_\ast/\epsilon_\ast)^{1/4}\simeq 0.025$, $n_{s,\ast} \simeq 0.9645$ and $r_\ast\simeq 0.002$, respectively. 

A model of large field inflation from this class is given by the potential 
\begin{align}
V(\phi) ~=~\,\, & 4.25907\cdot10^{-10} - 1.33159\cdot10^{-12} \phi \nonumber \\ 
& \hspace{-0.3cm} + \, 1.77313\cdot10^{-11} \phi^2 -  7.28605\cdot10^{-13} \phi^3 \nonumber \\
&\hspace{-0.3cm}+ \,e^{-1.2\, \phi} \big{(}1.1641\cdot10^{-13}- 1.05012\cdot10^{-9} \phi\nonumber \\
& \hspace{-0.3cm} + \, 2.76609\cdot10^{-11} \phi^2 + 4.67054\cdot10^{-13} \phi^3 \nonumber \\
& \hspace{-0.3cm} - \, 3.04863\cdot10^{-16} \phi^4\big{)}~.\label{exex2}
\end{align}
\begin{figure}[!h]%
    \hspace{-0.2cm}
    \includegraphics[width=8.7cm]{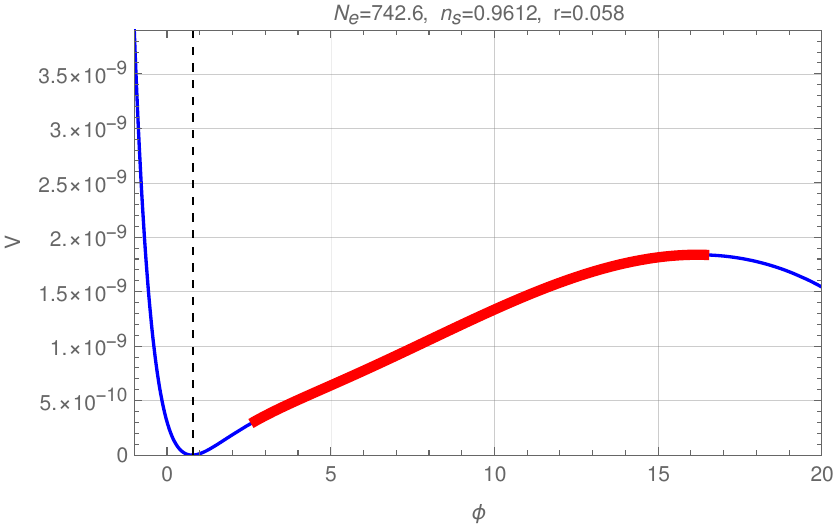} %
    \caption{\small 
    The quartic + exp $\times$ quartic potential from Eq.~\eqref{exex2}: large field inflation.}%
\label{fig:PolynExp2Ex2}
\end{figure}
The relevant parameters in this model are: $\phi_m\simeq 0.8$, $\phi_\ast\simeq 11.0$, $N_\ast\simeq 54.4$, $m=(V_\ast/\epsilon_\ast)^{1/4}\simeq 0.025$, $n_s\simeq 0.9613$, $r\simeq 0.058$. The model allows for a maximum number of $743$ e-folds.
\vspace{8pt}

\section{Other search methods}
\subsection{Reinforcement learning}
It is possible to interpret the space $\{s\}$ of inflationary models as an environment for the purpose of reinforcement learning. To this end we consider actions $a:s\mapsto s'$, which map a state $s$ to a new state $s'$ by changing the value of one of the coefficients in $s$ to the next smaller or larger one while leaving all other coefficients unchanged. The reward $R(s,a)$ of such an action can, for example, be defined as
\begin{equation}
 R(s,a)=\left\{\begin{array}{ccl}f(s')-f(s)&\mbox{for}&f(s')-f(s)\geq 0\\R_{\rm offset}&\mbox{for}&f(s')-f(s)<0\end{array}\right. \; ,
\end{equation} 
where $f$ is the fitness function for states defined previously and $R_{\rm offset}$ is a fixed (negative) value which penalises a decrease in fitness. This environment can then be coupled to one of the standard policy-based RL algorithms, such as REINFORCE or actor-critic (for a review see Refs.~\cite{sutton2018reinforcement,RUEHLE20201}).

Unfortunately, we find it difficult to train a neural network for such an inflationary environment, using either algorithm.  After careful adjustment of the fitness function we have managed to train successfully for a two-dimensional toy environment, consisting of potentials with just a quadratic and quartic term. We have not been able to achieve a successful training on any of the larger environments which were so efficiently handled by GAs. Intuitively, we attribute this failure to the ``choppiness" of the fitness function which can arise, for example, due to new global minima developing elsewhere in the potential, even for small parameter changes. Such discontinuities, clearly easily handled by GAs, can lead to unintended incentives for RL. The successful run for the two-dimensional toy environment suggests it is still conceivable that RL could be made to work on larger inflationary environments (see also the approach of Ref.~\cite{Rudelius:2018yqi} where gradient ascent in the number of e-folds was used to optimise the coefficients). This would probably require careful adjustment of the fitness function and systematic hyper-parameter optimisation but, given the efficiency of GAs, we have not attempted this. 

\subsection{Genetic programming}\label{sec:GP}
So far we have considered potentials of certain fixed functional forms, focusing on the optimisation of the numerical constants that can generate an inflationary potential. However in practice one may not know the functional form of the potential, but only the kinds of terms that can enter, for example polynomials for perturbative contributions,  trigonometric functions for axionic potentials, exponentials for nonperturbative processes and so forth. The way in which these components appear may vary depending on the physical set-up. Thus it is interesting to consider a more generic approach that can simultaneously optimise for both the functional form and the numerical factors. 

\begin{figure}[h]
    \centering
    \includegraphics[width=2cm]{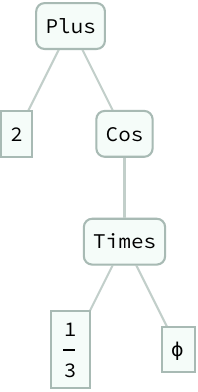} 
    \caption{\small Tree form of the potential $V(\phi)=2+\cos (\phi/3)$.}
\label{fig:treesimple}
\end{figure}

\begin{figure*}[t]
    \centering
    \includegraphics[width=18cm]{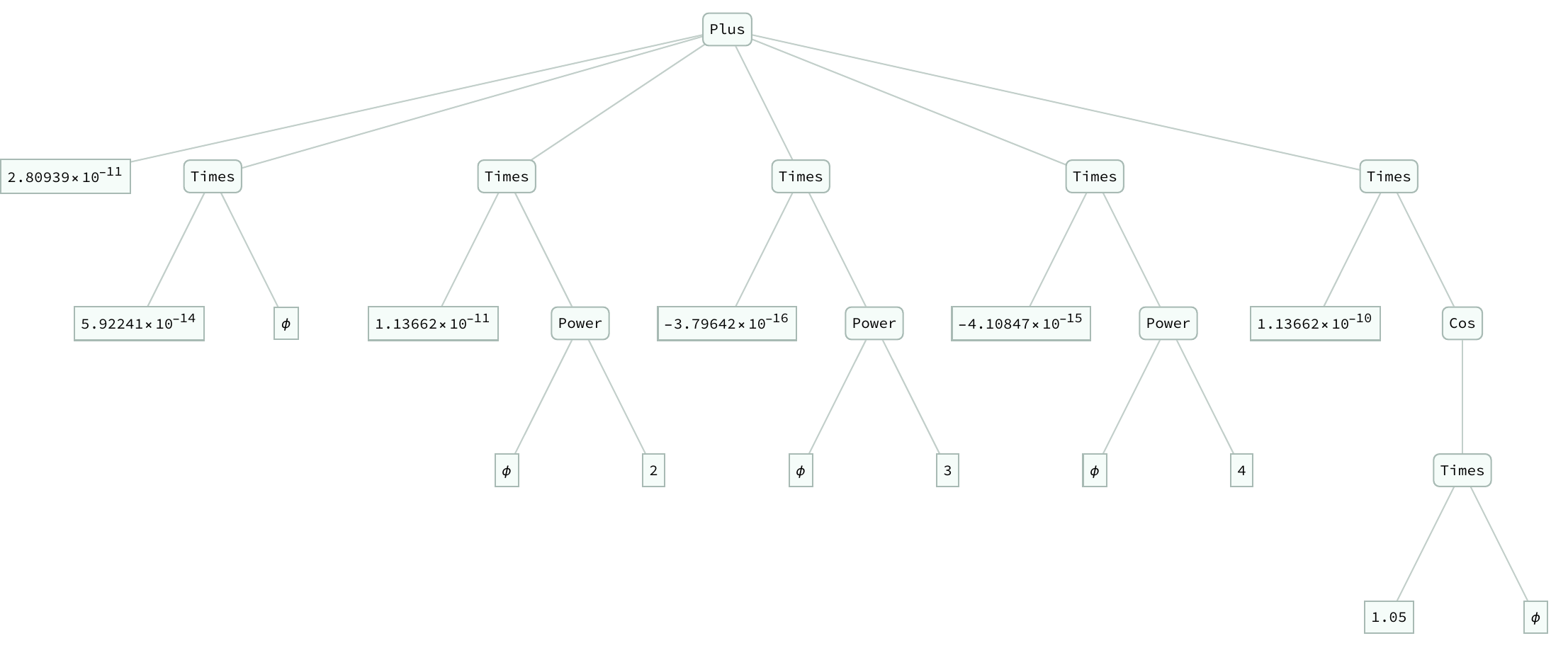} 
    \vspace{-1cm}
    \caption{\small Tree form of the potential in Eq.~\eqref{cosex}.}
\label{fig:tree}
\end{figure*}

For this purpose the paradigm of genetic programming (GP) is the  most promising setting. GP operates with trees of symbolic expressions and numerical factors. For example the tree form of the simple cosine potential $V(\phi)=2 + \cos(\phi/3)$ is shown in \fref{fig:treesimple}, and the tree for the inflationary cosine potential in Eq.~\ref{cosex}
is shown in~\fref{fig:tree}.

GP is an evolutionary algorithm that operates on populations of such trees and it shares with GAs the essential features of selection, crossover and mutation~\cite{koza:book,Koza89,banzhaf:1997:book,Poli2008}. It has been used successfully for many different tasks including for example symbolic regression (for a recent review that includes benchmark problems see Ref.~\cite{Cava2021}). Ideally one would operate to produce  a successful tree such as that in Figure \ref{fig:tree} as follows. Exactly as in a GA one first generates a random population, but in this case a population of trees, with the vertices being selected from a set of allowed operands, for example $\{ $~{\tt Times, Plus, Power, Cos}~$\}$, and the leaves being either constants or the variable $\phi$. Every tree can be ranked and selected for breeding as described already for the GA. However the process for breeding follows various well defined rules, with for example crossover implemented as a swap of subtrees (see Ref.~\cite{Poli2008} for a review). These can take many more forms than a simple one-point exchange of subtrees (so-called homologous cross-over), with subtrees possibly being moved to entirely different positions in the tree. After breeding, there is traditionally also a mutation stage. The effect of mutation is less clear cut in GPs and indeed it was even omitted from Koza's original work~\cite{Koza89}, however typically it {\it is} implemented and again can take many different forms.  For instance, mutation can affect individual leaves or entire subtrees (that is, a subtree could be replaced by a random subtree of comparable complexity). With these rules in place, the  evolutionary process then adheres to the same flow-chart as for the GA. 

Considering the case at hand, our experience with GAs indicates that it is crucial first to be clear about the question one would be trying to ask in this expanded framework. In particular if the form of the functions is incorporated into the search then the search space becomes virtually infinite. As we have seen, a suitable choice of coefficients can, for different functional forms, already yield an inflationary potential, so without specifying very carefully within the GP what we actually wish to achieve, the form of most potentials will be very inelegant. One can see the potential problems that can arise if we consider the more controlled example of the regression of a set of data that one already knows the answer to: for example the symbolic regression of data produced by our toy function $V(\phi)=2+\cos (\phi/3)$ (more precisely a set of fake data produced from say 100 random choices of $\phi$). We can ask the GP to fit this data, by for example defining the fitness function to be a monotonically decreasing function of the quadratic loss-function. However if we allow coefficients to be defined as we have been doing with 64 logarithmically spaced possible increments, then we quickly find that the algorithm runs into a severe version of the ``bloating"  phenomenon, in which expressions grow to progressively larger complexities. 

This suggests that when it comes to finding inflationary potentials, the precise question we should be asking is, given a particular set of operators that we know can be motivated by underlying physics, for example the aforementioned $\{ $~{\tt Times, Plus, Power, Cos}~$\}$ operands, what is the {\it simplest}~ form of potential that can produce inflation? 

In principle imposing simplicity on our expressions can reduce bloat. One can for example insert a fitness penalty that increases with the tree-depth. 
This is typically done in symbolic regression in order to avoid the GP equivalent of over-fitting. In the case of inflationary model building this trade-off would ideally render potentials of relatively low complexity, in line with the general preference towards finding simple underlying theoretical justifications, which can in addition correctly recover the allowed values for the cosmological observables.

Unfortunately the results for fitting the toy potential from its data are not encouraging. It is not possible to reduce bloat by simply adding a complexity penalty if one wishes to allow almost free-floating constants to exist within the GP. This is a well known issue when there are several constants to be determined, due to the functional form of the GP converging faster than the constants (see for example Ref~\cite{he2022taylor} for a related discussion). In the case of symbolic regression it is possible to circumvent this problem if the constants can instead be optimised independently by for example nonlinear regression. 
In the present example, one would in practice allow the numerical coefficients to be undefined parameters in the tree, with the corresponding leaves filled with `placeholders'. In order to work out the fitness of any particular tree in the population, one would then  determine the set of coefficients that separately maximises the fitness function in a similar fashion to that described in Refs.~\cite{Krawiec2002,Korns2011,Worm2013}.
Indeed we find that this method works well in our toy symbolic regression problem. 

The present problem is, as we have stressed, more general than a symbolic regression, however the examples above  suggest ways forwards. Implementing GPs to find optimal functions should be organised in a nested fashion, similar to a technique that has been called {\it speciation} in the literature~\cite{Duda1996,10.1115/1.2828774,MARTINS2020113191}, and a related technique called {\it niching}. The idea of speciation as it would be applied to the inflationary potential problem is as follows. The evolutionary algorithm is arranged in two stages. The first is the GP, organised as described above, with any constants in the tree represented by the aforementioned placeholders. Then, in order to calculate the fitness of all the trees in a generation, one has to run a GA on each individual function in the population (which effectively defines a sub-population within a {\it niche}) to optimise its fitness,  following the procedure described in the main body of this paper. This simultaneously determines the values of all the constants for each function.

The above procedure is guaranteed to work (modulo the need to implement good ranges for the constants), because we know each half of the procedure works separately, so in that sense it constitutes a ``no-lose theorem". However it could clearly end up being  computationally expensive. If the GP part of the problem requires a population of $N_{GP}$ niches, then one might expect to incur a cost of a factor of at least $N_{GP}$ compared to the GA. One could possibly reduce the penalty if the GP is still ``niched'', but with each niche containing fewer members than the individual GA, the rationale being that niching merely needs to speed up the convergence of constants enough to overcome the tendency to bloat.  We shall return to these more ambitious approaches in a future publication. 

\section{Conclusions}
This note has expanded the range of computational tools available in inflationary model building to include genetic algorithms. We have shown that GAs can be efficiently used to identify models of inflation that are consistent with the required number of e-folds, the current bounds on the spectral index of scalar perturbations, the tensor-to-scalar ratio, and the scale of inflation. Using moderate computational resources we have been able to construct millions of viable models, however many more can be obtained in this way. 

In particular, we have focused on three functional classes of potentials: (1) sextic polynomials, (2) quartic polynomials + cosine and (3) quartic + exp $\times$ quartic. In the case of pure polynomial potentials we attempted a semi-comprehensive search which resulted in the interesting observation that low-$r$ models are favoured. 

Reinforcement learning, which in other contexts has been shown to perform comparably well~\cite{Abel:2021rrj,Abel:2021ddu,Cole:2021nnt}, was difficult to use for the present problem. On the other hand, methods involving genetic programming and symbolic regression seem promising approaches to explore. They may lead to to viable potentials with new functional forms and could provide fresh inspiration for model building. \\

\section*{Acknowledgements}
AC's research is supported by a Stephen Hawking Fellowship (EPSRC grant $\text{EP/T016280/1}$). The research activities of SAA were supported by the STFC under grant $\text{ST/P001246/1}$ and the Institut Pascal at
Universit\'e Paris-Saclay. TRH is supported by an STFC studentship. TRH and AL would like to thank Aditi Chandra for a related collaboration early in this project.

\vspace{30pt}

\bibliography{bibliography}
\bibliographystyle{inspire}

\end{document}